# Electric Field-Induced Nonreciprocal Directional Dichroism in a Time-Reversal-Odd Antiferromagnet

Takeshi Hayashida[*,†], Koei Matsumoto[*], Tsuyoshi Kimura
*Department of Applied Physics, University of Tokyo, Bunkyo-ku, Tokyo 113-8656, Japan*

Antiferromagnets with broken time-reversal ($\mathcal{T}$) symmetry ($\mathcal{T}$-odd antiferromagnets) have gained extensive attention, mainly due to their ferromagnet-like behavior despite the absence of net magnetization. However, certain types of $\mathcal{T}$-odd antiferromagnets remain inaccessible by the typical ferromagnet-like phenomena (e.g., anomalous Hall effect). One such system is characterized by a $\mathcal{T}$-odd scalar quantity, the magnetic toroidal monopole. To access the broken $\mathcal{T}$ symmetry in such a system, we employ a unique nonreciprocal optical phenomenon, electric field-induced nonreciprocal directional dichroism ($E$-induced NDD). We successfully detected signals of $E$-induced NDD in a $\mathcal{T}$-odd antiferromagnet, $Co_2SiO_4$, whose magnetic structure is characterized by the magnetic toroidal monopole. Furthermore, by spatially resolving the $E$-induced NDD, we visualized spatial distributions of a pair of domain states related to one another by the $\mathcal{T}$ operation. The domain imaging revealed the inversion of the domain pattern by applying a magnetic field, which is explained by trilinear coupling attributed to the piezomagnetic effect. Our observation of $E$-induced NDD provides a unique approach to accessing the order parameter in $\mathcal{T}$-odd antiferromagnets.

## I. INTRODUCTION

Antiferromagnetism is a state in which magnetic moments are ordered in a way that the overall magnetization is canceled out. Although Louis Néel [1], who was awarded the Nobel Prize for research on antiferromagnetism in 1970, described it as "interesting but useless," a considerable amount of research since his comment has revealed that certain antiferromagnets exhibit unique functionalities or physical phenomena. These properties often arise from the breaking of time-reversal ($\mathcal{T}$) symmetry. For example, in an insulating antiferromagnet where space-inversion ($\mathcal{P}$) and $\mathcal{T}$ symmetries are simultaneously broken but the combined $\mathcal{PT}$ symmetry is preserved, the linear magnetoelectric (ME) effect emerges [2–4]. Such ME antiferromagnets also exhibit unique nonreciprocal optical and transport properties [5]. For another example, recent investigations have revealed that antiferromagnets breaking $\mathcal{T}$ and $\mathcal{PT}$ symmetries show physical properties that are usually observed in ferromagnets, including the anomalous Hall effect, spontaneous magneto-optical Kerr effect, and nonrelativistic spin splitting, despite the absence of (or tiny) net magnetization [6–13]. To understand such unconventional properties in $\mathcal{T}$- and $\mathcal{PT}$-odd antiferromagnets with collinear and compensated magnetism, a novel concept termed "altermagnetism" has been introduced [14,15]. Notably, altermagnetism has attracted extensive interest partly because of the above-mentioned ferromagnet-like properties.

Herein, we examine another unique property of $\mathcal{T}$- and $\mathcal{PT}$-odd antiferromagnets, electric field-induced ($E$-induced) nonreciprocal directional dichroism (NDD), where the application of $E$ induces asymmetry in optical absorption between two counterpropagating light beams. Using this unique nonreciprocal optical phenomenon, we successfully detect the order parameter of a $\mathcal{T}$- and $\mathcal{PT}$-odd antiferromagnet and access its domain states.

## II. ELECTROTOROIDIC EFFECT AND ELECTRIC FIELD-INDCED NONRECIPROCAL DIRETIONAL DICHROISM

### A. Electrotoroidic effect

The targeted class of materials in this research is an antiferromagnet which breaks $\mathcal{T}$, but the above-mentioned ferromagnet-like behaviors are symmetrically forbidden. Even in such materials, however, $\mathcal{T}$-odd property still hosts interesting physical phenomena. One such phenomenon is the diagonal electrotoroidic (ET) effect [16,17], where the application of an electric field **E** induces magnetic toroidal moment **T** in the direction parallel to **E**. **T** is a $\mathcal{T}$-odd polar vector that describes vortex arrangements of magnetic dipoles as $\mathbf{T} \propto \sum_i \mathbf{r}_i \times \mathbf{m}_i$, where $\mathbf{r}_i$ is the position vector of the magnetic dipole $\mathbf{m}_i$ at the $i$ site with reference to the high symmetry point (middle panel of Fig. 1). The spontaneous ferroic order of magnetic toroidal moments, so-called ferrotoroidic order, has long been discussed theoretically [16,18–20], and ferrotoroidic domains were experimentally observed using optical second harmonic generation in 2007 [21]. In parallel with the discussion on the ferrotoroidic order, Schmid introduced a linear response of **T** to applied **E**, described as $T_i = \theta_{ij}E_j$ and termed it the ET effect [16,17]. Here, $\theta_{ij}$ is a second rank $\mathcal{T}$-odd polar tensor called the ET tensor. Note that the ET effect is a phenomenon ascribed to a bi-linear ME coupling since **T** is symmetrically equivalent to **H**×**E**, where **H** is a magnetic field, and the coupling between **T** and **E** is expressed as (**H**×**E**)·**E** [16,22]. When $i \neq j$ and $i = j$, the effect is off-diagonal (**E**⊥**T**) and diagonal (**E**||**T**), respectively. Off-diagonal coupling is rather common and is observed in systems breaking both $\mathcal{P}$ and $\mathcal{T}$ symmetries with the cross product of spontaneous (or field-induced) **M** and an applied **E**, i.e., **T** ∝ **M**×**E** (right panel of Fig. 1). The off-diagonal ET effect has been observed through nonreciprocal optical phenomena in systems with **E** and **H** applied in perpendicular directions (**T** ∝ **H**×**E**) [23].

[*]These authors contributed equally to this work.
[†]Corresponding author: thayashida@ap.t.u-tokyo.ac.jp

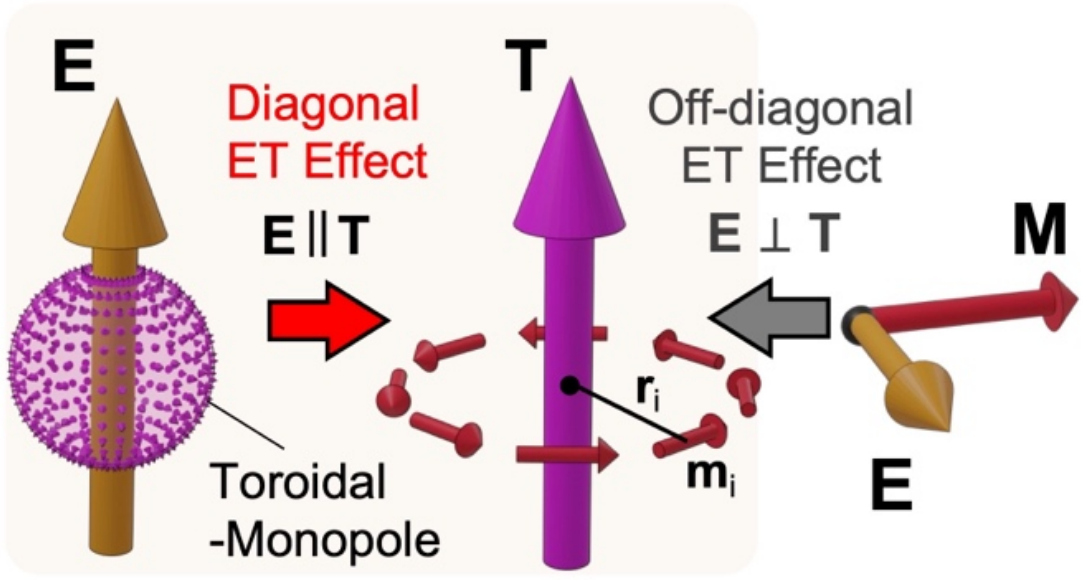

FIG. 1. Magnetic toroidal moment and electrotoroidic (ET) effect. Magnetic toroidal moment **T** is composed of vortex arrangements of magnetic dipoles $\mathbf{m}_i$ (middle panel). **T** can be induced by an electric field **E** as a result of the ET effect. The off-diagonal ET effect refers to **T** induced perpendicular to **E** (**E**⊥**T**), which is achieved by a cross-product of magnetization **M** and an applied **E**, **T**∝**M**×**E** (right panel). The diagonal ET effect, **T** induced parallel to **E** (**E**||**T**), is mediated by a $\mathcal{T}$-odd scalar quantity, such as magnetic toroidal monopole (left panel).

On the other hand, for the emergence of the diagonal ET effect, **T** is induced as a result of changing only the $\mathcal{T}$ symmetry ($\mathcal{T}$-even to $\mathcal{T}$-odd) while maintaining the direction of **E**, which should be observed in systems holding a $\mathcal{T}$-odd "scalar" quantity. Recently, Hayami and Kusunose showed that such a class is characterized by "magnetic toroidal monopole", that is, all-in or all-out arrangements of the local magnetic toroidal moment [24] (left panel of Fig. 1). Considering symmetry, 32 magnetic point groups allow for the emergence of the magnetic toroidal monopole and therefore the diagonal ET effect [24]. More recently, a similar classification focusing on spontaneous and field-induced **T** has been introduced by Xu and coworkers [25]. We note that the systems described by the magnetic toroidal monopole can also be described by the magnetic octupole [26,27]. The magnetic toroidal monopole better reflects $\mathcal{T}$-odd scalar quantity that we focus on in this research, and thus we adopt the classification using the magnetic toroidal monopole. We also note that the bi-linear ME coupling is suggested as an effective but unexplored way to approach magnetic octupole orders [28].

### B. Electric field-induced nonreciprocal directional dichroism

In contrast with the measurement of **M**, that of **T** is non-trivial, and thus methods to observe the diagonal ET effect have been less established. However, in systems with finite **T**, the characteristic optical phenomena, NDD, emerges. NDD has been observed in ferrotoroidic materials and materials in which magnetic toroidal moment is induced by the cross product of **E** (**P**) and **H** (**M**) [29–32]. Because the states with opposite sign of **T**, T+ and T–, show different absorption, NDD has been used to visualize ferrotoroidic domains [33,34]. In the system where the diagonal ET effect, i.e., $E$-induced **T**, is allowed, therefore, the emergence of $E$-induced NDD is expected.

$E$-induced NDD will be described as the changes in an absorption coefficient $\alpha$ [cm$^{-1}$] under **E**, and expressed as

$$\alpha = \alpha_0 + \Delta\alpha = \alpha_0 + \beta(\mathbf{k}\cdot\mathbf{E}), \tag{1}$$

where $\alpha_0$ is an absorption coefficient without **E**, $\Delta\alpha$ is the difference in the absorption coefficient induced by **E**, $\beta$ is the coefficient describing the $E$-induced NDD, and **k** is the light propagation vector. When **E** is parallel to **k**, the intensity of the transmitted light $I$ is described as

$$I = I_0 e^{-(\alpha_0+\Delta\alpha)d} \approx I_0 e^{-\alpha_0 d}(1-\Delta\alpha d)$$
$$= I_0 e^{-\alpha_0 d}(1-\beta V), \tag{2}$$

where $I_0$ is the intensity of the incident light, $d$ [cm] is the sample thickness, and $V = Ed$ [V] is the applied voltage. In the transformation of Eq. (2), $\Delta\alpha d$ is assumed to be small. Under this approximation, the change in transmitted light intensity is linear relative to $V$. Note that this $E$-induced NDD is understood as the coupling of ($\mathbf{H}_\omega\times\mathbf{E}_\omega$)·**E**, where $\mathbf{E}_\omega$ and $\mathbf{H}_\omega$ are oscillating electric field and magnetic field of light, respectively. Thus, $E$-induced NDD is also a type of bi-linear (optical) ME coupling.

### C. Time-reversal-symmetry-broken antiferromagnet $Co_2SiO_4$

To manifest the $E$-induced NDD, we selected $Co_2SiO_4$ as the target material. $Co_2SiO_4$ crystalizes in the olivine structure [space group *Pnma*, Fig. 2(a)] [35–37]. $Co^{2+}$ ions are surrounded by six $O^{2-}$ ions and occupy two different octahedral sites: Co1 located at the centrosymmetric site (site symmetry $\bar{1}$) and Co2 located at the mirror-symmetric site whose mirror plane is perpendicular to the $b$ axis (site symmetry $m$). $Co_2SiO_4$ shows an antiferromagnetic (AFM) transition at $T_N \approx 50$ K [38–40]. Figure 2(c) shows the temperature profile of magnetic susceptibility of our $Co_2SiO_4$ crystal. In the AFM phase, spins on the Co2 sites are collinearly aligned along the *b* axis, whereas those on the Co1 sites are slightly tilted from the *b* axis and form zigzag spin chains [red arrows in Fig. 2(a)]. The magnetic point group is *mmm*, which holds the magnetic toroidal monopole [24] and permits the diagonal ET effect with the nonzero distinct ET tensor components of $\theta_{11}$, $\theta_{22}$, and $\theta_{33}$ [41]. Although the noncollinear spins on the Co1 sites are also related to the breaking of $\mathcal{T}$ symmetry [25], herein, we focused on the collinear spins on the Co2 site for simplification. As mentioned above, **T** describes vortex arrangements of multiple magnetic dipoles; however, local magnetic toroidal moment **t** can also be defined at a single ion site by considering the cross product of a local electric dipole moment **p** and a local magnetic moment **m**, as $\mathbf{t} \propto \mathbf{p}\times\mathbf{m}$ [32]. At the Co2 site with $\mathbf{m}_{Co2}$ [red arrows in Fig. 2(b)], an inversion symmetry is locally broken while a mirror symmetry perpendicular to the *b* axis is preserved, and thus $\mathbf{p}_{Co2}$ emerges in the *ac* plane [yellow arrows in Fig. 2(b)]. Then, a local magnetic toroidal moment at the Co2 site, $\mathbf{t}_{Co2}$ ($\propto \mathbf{p}_{Co2}\times\mathbf{m}_{Co2}$) is introduced [magenta arrows in Fig. 2(b)]. Note that the sum of $\mathbf{t}_{Co2}$ is canceled out in a unit cell with four Co2 sites; therefore, $Co_2SiO_4$ is not ferrotoroidic ($\mathbf{T} = \mathbf{0}$) in the absence of **E**. We also note that spin-up and spin-down sublattices are related to each other by glide operations but not by space-inversion or translation, which satisfies the requirement of altermagnetism from the viewpoint of symmetry [42,43], i.e., $Co_2SiO_4$ is an altermagnet candidate.

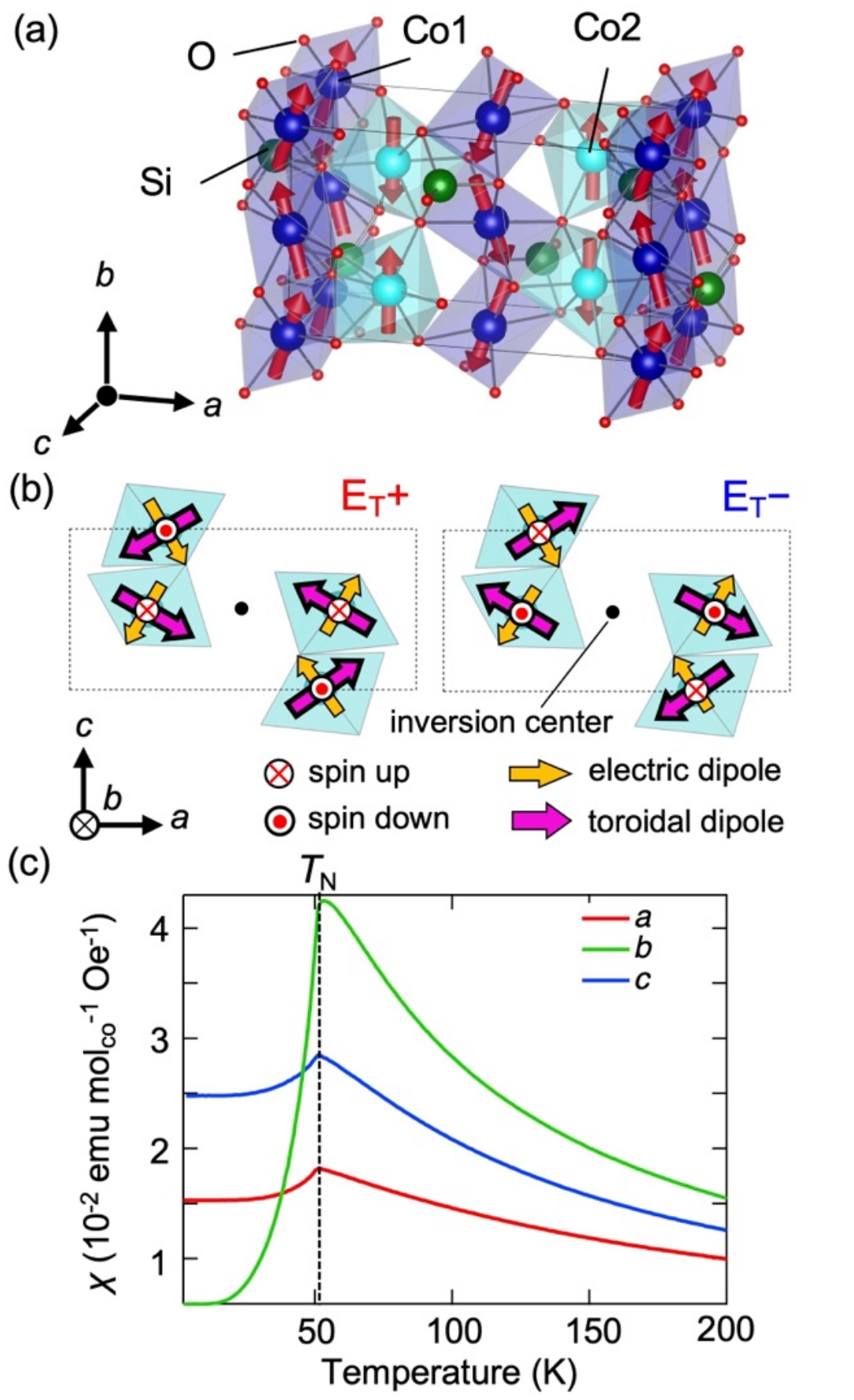

FIG. 2. (a) Crystal and magnetic structures of $Co_2SiO_4$. The red arrows denote $Co^{2+}$ spins. (b) Two magnetic domain states ($E_T$+ and $E_T$–) related to one another via time-reversal operation. For clarity, only $CoO_6$ octahedral units at the Co2 sites are shown. The circles with red dots or crosses in the center, yellow arrows, and magenta arrows denote local spins, electric dipoles, and magnetic toroidal dipoles, respectively. The $E_T$+ and $E_T$− domain states are characterized by magnetic toroidal monopoles with a positive ($\sum_i \mathbf{r}_i \cdot \mathbf{t}_i > \mathbf{0}$) and negative ($\sum_i \mathbf{r}_i \cdot \mathbf{t}_i < \mathbf{0}$) divergence of $\mathbf{t}$, respectively. The dashed box denotes a unit cell. (c) Temperature profiles of magnetic susceptibility $\chi$ along the *a*, *b*, and *c* axes.

In the absence of **E**, thus, the AFM phase of $Co_2SiO_4$ does not show finite **T** but holds a $\mathcal{T}$-odd scalar quantity, that is, magnetic toroidal monopole ($\propto \sum_i \mathbf{r}_i \cdot \mathbf{t}_i$) [24]. One can consider the two domain states related to each other by the $\mathcal{T}$ operation, termed $E_T$+ and $E_T$– domains [(left and right panels of Fig. 2(b)], which are characterized by the magnetic toroidal monopole with positive ($\sum_i \mathbf{r}_i \cdot \mathbf{t}_i > 0$) and negative ($\sum_i \mathbf{r}_i \cdot \mathbf{t}_i < 0$) divergence, respectively. The sign of the ET tensors of the two $E_T$ domain states are opposite, indicating that these states exhibit **T** with opposite polarities (T+ and T–) under **E**. This results in the opposite sign of $\beta$ in Eq. (2), i.e., the two domain states show different absorption under **E**. Figure 3 schematically illustrates the ET effect and the $E$-induced NDD expected in a multidomain state of $Co_2SiO_4$. The way to align the system into a single domain state is nontrivial. Therefore, spatially resolved measurements of $E$-induced NDD are required to detect the ET effect.

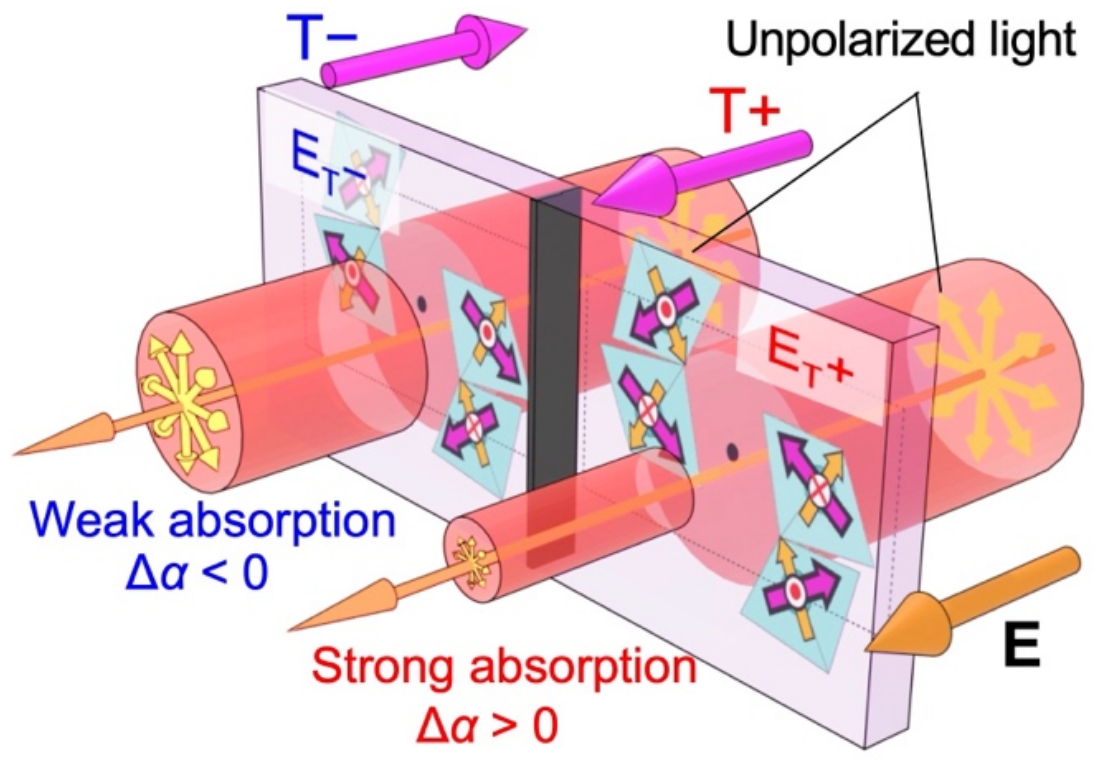

FIG. 3. *E*-induced NDD in $Co_2SiO_4$. Under an electric field **E**, the $E_T$+ and $E_T$– domain states exhibit magnetic toroidal moments with the opposite polarities (T+ and T−), resulting in an absorption difference of unpolarized light. Although here we depict the model in the setting of $\mathbf{E}\|\mathbf{k}\|b$, *E*-induced NDD is permitted in any direction if $\mathbf{E}\|\mathbf{k}$ is satisfied.

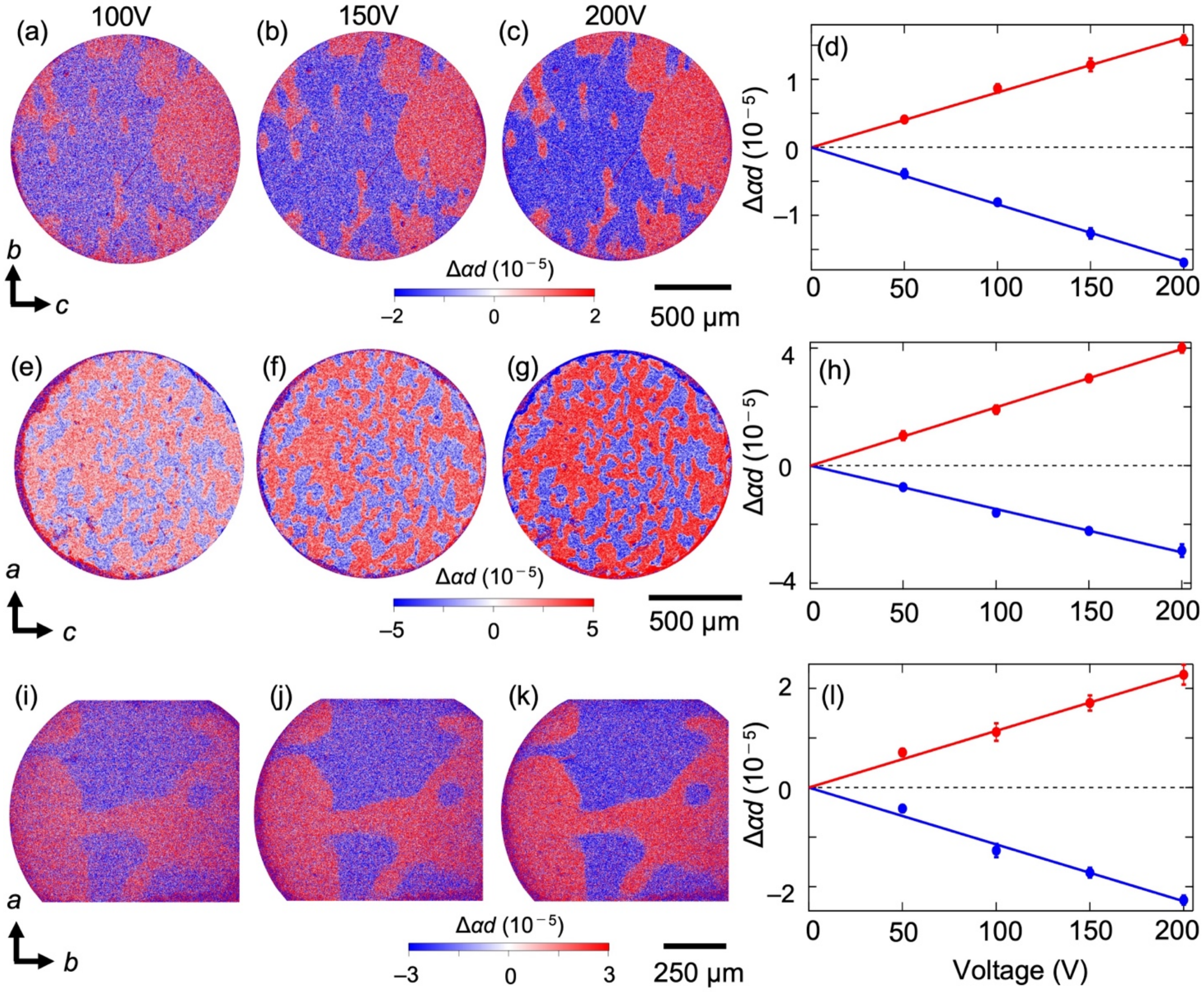

FIG. 4. $E_T$ domains observed by $E$-induced NDD. (a-c), (e-g), and (i-k) show maps of changes in the absorption of unpolarized light ($\Delta\alpha d$) induced by applying voltage in the (100), (010), and (001) planes, respectively. The first (a,e,i), second (b,f,j), and third (c,g,k) columns show the images obtained under applied voltages of 100, 150, and 200 V. respectively. The applied voltage and light propagation direction **k** were parallel to the normal axis of each plane. The wavelengths of the incident light were 550 nm (a-c and i-k) and 590 nm (e-g). The images were obtained at 4 K. (d), (h), and (l) show the voltage dependence of the $\Delta\alpha d$ averaged in 10 different points for each domain state in the (100), (010), and (001) planes, respectively. The sampling positions are shown in Fig. 8 in Appendix B. The red and blue dots correspond to the data for the positive and negative $\Delta\alpha d$ single domain areas, respectively. The error bars show the standard deviations for the 10 points. The lines denote the least-squares fitting lines.

## III. RESULTS AND DISCUSSION

### A. Observation of *E*-induced NDD

We measured $E$-induced NDD of $Co_2SiO_4$ single crystals. For the measurements in the transmittance geometry, we prepared three thin samples whose widest faces are parallel to the (100), (010), and (001) planes (see Appendix A.1). Based on other $E$-induced optical effects in magnetic materials (e.g., $E$-induced magnetic circular dichroism in $Cr_2O_3$ [44]), $E$-induced NDD signals are expected to be weak. Thus, we adopted an electric-field-modulation imaging technique [45], which enables us to detect the spatial distribution of $E$-induced NDD signals with high sensitivity (see Appendix A.2). Figures 4(a)-4(c), 4(e)-4(g), and 4(i)-4(k) show the maps of $\Delta\alpha d$ obtained in the geometry where both **k** and **E** are along the $a$, $b$, and $c$ axes, respectively, at selected $V$. The images were obtained at 4 K (< $T_N$) using unpolarized light with wavelengths of 550 nm [Figs. 4(a)-4(c) and 4(i)-4(k)] and 590 nm [Figs. 4(e)-4(g)]. In all the samples, a clear contrast of red and blue corresponding to the positive and negative $\Delta\alpha$, respectively, is observed at 200 V [Figs. 4(c),4(g), and 4(k)]. The color contrasts monotonically increase with voltage in all the geometries. We calculated the average $\Delta\alpha d$ for the pixels at selected single domain areas (both positive and negative $\Delta\alpha d$) denoted by boxes in Figs. 8(a)-8(c) of Appendix B and determined the $V$ dependence of $\Delta\alpha d$. As shown in Figs. 4(d), 4(h), and 4(l), $\Delta\alpha d$ is linear with respect to $V$. We also checked that the contrasts disappeared at temperatures above $T_N$ (Fig. 5). Furthermore, the lock-in measurements with a focused laser beam in a detailed temperature range (see Appendix A.3) confirm that the $E$-induced NDD disappears above $T_N$ (Fig. 9 in Appendix C). Consequently, the obtained contrasts are pairs of $E_T$ domains showing the opposite sign of $E$-induced NDD resulting from $\mathcal{T}$ symmetry breaking. The effect is observed in all the settings of **E**||**k**||$a$, $b$, $c$, which is consistent with the magnetic point group $mmm$ where all the diagonal ET tensor components ($\theta_{11}$, $\theta_{22}$, and $\theta_{33}$) can be finite. Note that the signs of the domain states between the different planes cannot be uniquely associated in the present measurements; thus, the color (the sign of $\Delta\alpha$) coincidence of the domains between Figs. 4(c), 4(g), and 4(k) does not necessarily indicate the coincidence of the domain states.

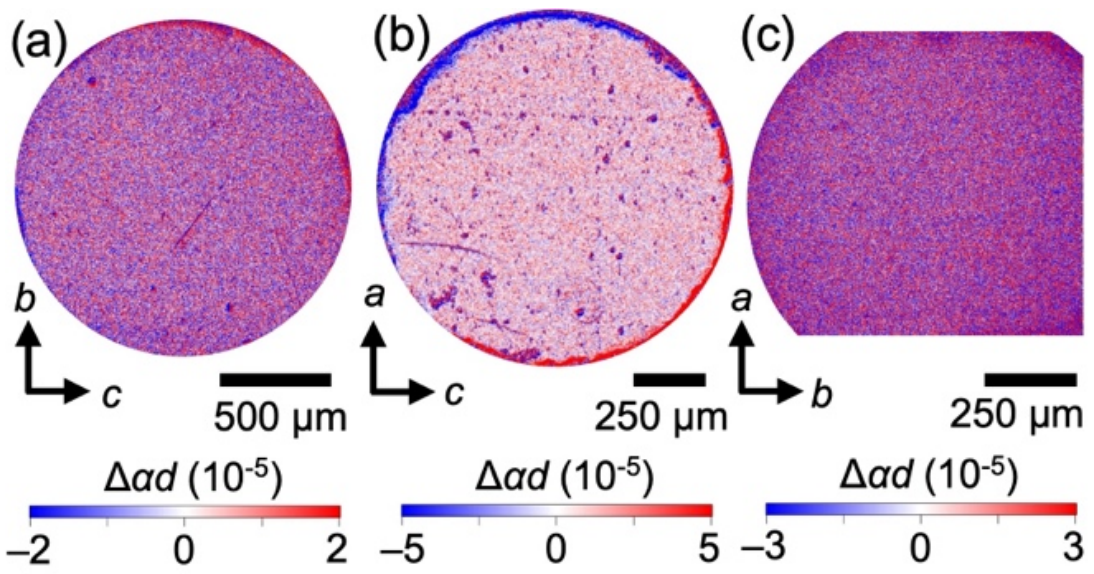

FIG. 5. Spatial maps $\Delta\alpha d$ above $T_N$. (a), (b), and (c) show the maps of changes in $\Delta\alpha d$ induced by an applied voltage of $V = 200$ V in the (100), (010), and (001) planes, respectively, at temperatures above $T_N = 50$ K [(a):52 K, (b):61 K, (c):53 K]. Applied electric field **E** and light propagation direction **k** are parallel to the normal axis of each plane. The wavelengths of the incident light are 550 nm (a and c) and 590 nm (b). The images in panels (a), (b), and (c) were taken at the same position on the same sample as those in Figs. 4(a)-4(c), 4(e)-4(g), and 4(i)-(k), respectively.

The domain sizes on the (100) and (001) planes [Figs. 4(c) and 4(k)] are on the order of several hundred micrometers, whereas that on the (010) plane [Fig. 4(g)] is on the order of several tens of micrometers. The domains on the (010) plane are expected to be vertically cut from the domains on the (100) and (001) faces along the $b$ axis, which does not match the experimental results. As discussed below, the domains of $Co_2SiO_4$ are probably related to strain, and the domain size may be affected by residual strains induced during sample preparation, particularly, cutting and polishing.

We have also acquired the spectra of $E$-induced NDD by focusing a laser beam on a single domain region and using a lock-in technique (see Appendix A.3). Although NDD itself does not depend on the polarization of incident light, polarization affects the absorption and $E$-induced NDD because of the orthorhombic crystal structure. Therefore, the measurements were performed using linearly polarized light. Complex peak structures of $E$-induced NDD were observed in the energy range the Co $d$-$d$ transition (Fig. 10 in Appendix D) [46]. For a detailed analysis and discussion of the spectra, see Appendix D.

### B. Domain inversion by applying a magnetic field

Because the magnetic point group $mmm$ of $Co_2SiO_4$ forbids spontaneous magnetization, the $E_T$ domains were expected to not respond to a magnetic field **H**. However, the domains exhibited an anomalous response to **H**. Figure 6 shows the domain maps on the (100) plane obtained at 4 K in the absence of **H** after cooling the samples across $T_N$ while applying **H** of ± 50 mT along the $c$ [Figs. 6(a) and 6(b)], $b$ [Figs. 6(c) and 6(d)] and $a$ [Figs. 6(e) and 6(f)] axes. The domain strongly depends on the direction of the cooling **H**. However, a single-domain structure has never been obtained in any **H** direction. The most surprising feature is that the domain contrasts are completely inverted by reversing the direction of cooling **H** while maintaining the positions of the domain boundaries [compare Fig. 6(a) with 6(b), Fig. 6(c) with 6(d), and Fig. 6(e) with 6(f)]. We also checked how the domains respond when **H** is applied below $T_N$ (Fig. 11 in Appendix E). Figure 11 shows the result for the (100) plane sample at 49 K (just below $T_N$) in **H**||$c$. The application of **H** changes the domain structure, and, again, the domain contrast was inverted depending on the sign of **H**.

This domain inversion in response to the flipping of the sign of **H** provides insights into the mechanism of the unexpected response of the $E_T$ domains. Similar domain inversion has been reported in some multiferroics (e.g., $Mn_2GeO_4$ and $Dy_{0.7}Tb_{0.3}FeO_3$) in which three order parameters (AFM order parameter $L$, magnetization $M$, and electric polarization $P$) are coupled via a trilinear coupling, and the Landau free energy is described as $F \propto -LMP$ [47–49]. When one of the order parameters, e.g., $L$, is fixed, the other two ($M$ and $P$) are coupled to minimize the free energy, and the domain switching of $M$ accompanies that of $P$. In this case, the domain pattern of $P$ is clamped by that of $L$, but the sign of $P$ in each domain is inverted in response to a sign reversal of $M$.

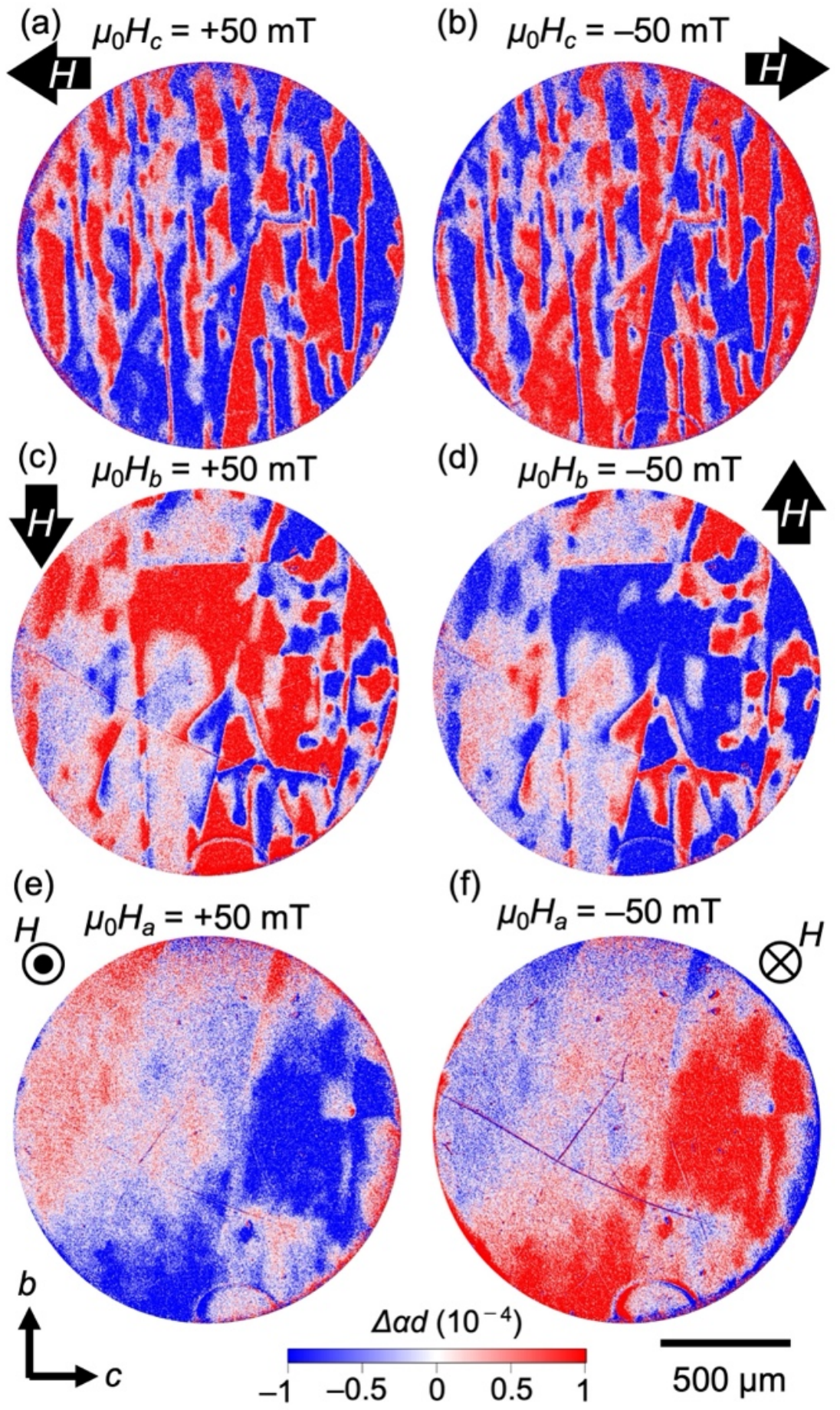

FIG. 6. $E_T$ domain inversion by applying a magnetic field**.** (a)-(f) $E_T$ domains in the (100) plane obtained at 4 K in the absence of a magnetic field after cooling the sample across $T_N$ in a magnetic field of ±50 mT along the $c$ axis (a,b), $b$ axis (c,d) and $a$ axis (e,f); $c$-polarized light with a wavelength of 590 nm was used for imaging, and the applied voltage was 200 V. The domains obtained by applying positive and negative magnetic fields show the same pattern, but the contrast is inverted [(a) vs. (b), (c) vs. (d), and (e) vs. (f)].

In the case of the $\mathcal{T}$-odd antiferromagnet $Co_2SiO_4$, $L$ is the only well-defined order parameter, which is equivalent to the order parameter of the $E_T$ domain. However, considering the observed response to **H**, magnetization can be another order parameter. In fact, our magnetization measurements revealed tiny spontaneous magnetization below $T_N$ (Fig. 12 in Appendix F). Although the observed spontaneous magnetization is very small (on the order of $10^{-4}$ $\mu_B$/f.u.), its emergence contradicts the magnetic point group of $Co_2SiO_4$.

Here we consider that the origin of magnetization is ascribed to a linear piezomagnetic effect, a linear coupling between magnetization and strain $\sigma$, which is permitted in $\mathcal{T}$-odd antiferromagnets characterized by the magnetic toroidal monopole or magnetic octupole [28]. In fact, the piezomagnetic effect is observed in several $\mathcal{T}$-odd antiferromagnets, including the spin reorientation phase of $DyFeO_3$ [50] whose magnetic point group is *mmm* and the same as $Co_2SiO_4$. The linear piezomagnetic effect is defined as $M_i = \Lambda_{ijk}\sigma_{jk}$ [51–53] where $\Lambda_{ijk}$ is a piezomagnetic tensor. In the magnetic point group *mmm*, $\Lambda_{abc}$, $\Lambda_{bca}$, and $\Lambda_{cab}$ are finite, and their sign depends on the $E_T$ domain state [41]. Thus, if there is an internal shear strain in $Co_2SiO_4$, a trilinear coupling described as $F = -M_i\sigma_{jk}L$ is expected. This strain may be induced by cutting or polishing the sample and/or intrinsic distortion of the crystal lattice. When shear-stain domains are spatially distributed and remain unchanged during the application of **H**, the observed inversion of $E_T$ domains caused by the application of **H** is well explained (Fig. 7). Because the application of **H** transforms the $M$ domain into a single domain state, the shear-strain and $E_T$ domains are clamped, and a reversal of $M$ leads to the inversion of $E_T$. The direction of **H** determines which shear-strain distributions respond ($\sigma_{ab}$ for $H_c$, $\sigma_{ca}$ for $H_b$, and $\sigma_{bc}$ for $H_a$). Thus, as observed in our sample, the $E_T$ domain structure depends on the direction of **H** [compare Figs. 6(a), 6(c) and 6(e)]. The presence of the strain domains in our samples was not confirmed experimentally, which will be for future work.

## IV. Conclusion

In conclusion, we investigated the $E$-induced NDD in the $\mathcal{T}$-odd antiferromagnet $Co_2SiO_4$. This effect is unique to materials characterized by the order of the $\mathcal{T}$-odd scalar quantity, such as the magnetic toroidal monopole. In such materials, the anomalous Hall effect and the spontaneous magneto-optical Kerr effect do not manifest, for which reason these materials have been less investigated. As proved in this study, the $E$-induced NDD is highly effective for revealing the order of the $\mathcal{T}$-odd scalar quantity, more particularly visualizing domains. The $E$-induced NDD occurs in an insulator to which an electric field can be applied. Until now, discussions on $\mathcal{T}$- and $\mathcal{PT}$-odd antiferromagnets, particularly those centered on altermagnets, have mainly focused on metallic systems [14,15]. Our research contributes to the extension of such studies to insulators. Candidate materials are $DyFeO_3$ [50] and $Mn_2GeO_4$ [47].

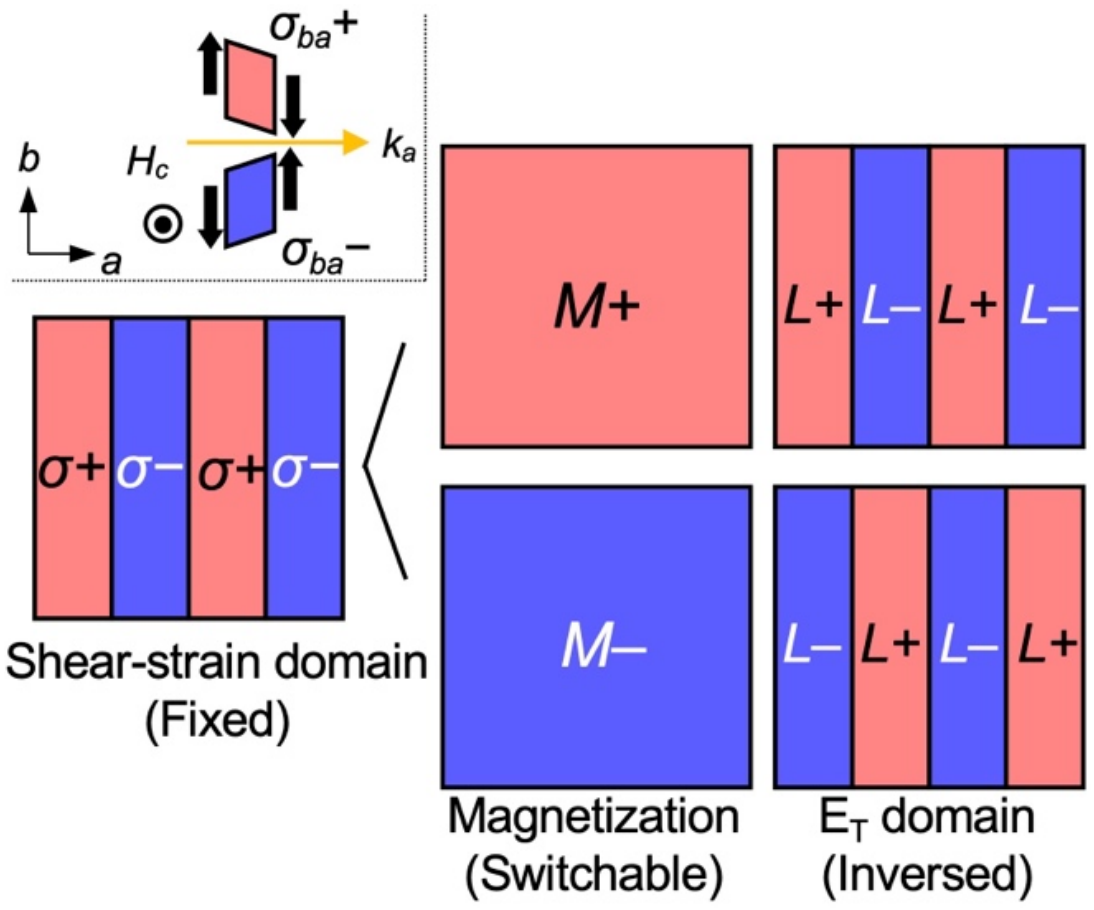

FIG. 7. $E_T$ domain inversion model with a trilinear coupling. Red and blue represent the positive and negative domain states for each order parameter (shear strain: $\sigma$, magnetization: $M$, $E_T$ domain: $L$), respectively. The shear-strain domain remains rigid and invariant under a magnetic field (left panel). Thus, when a magnetic field is applied and magnetization is uniform (M+ or M−), the $E_T$ domain pattern mirrors the shear-strain distribution. The sign of $E_T$ domain is inverted for M+ and M−, to satisfy the trilinear coupling of $F = -\sigma ML$. The inset figure shows the situation corresponding to Fig. 6(a), where the (100) plane domain is obtained with a magnetic field applied along the $c$ axis. In this case, the distribution of $\sigma_{ba}$ determines the $E_T$ domain structure.

The domain imaging using $E$-induced NDD revealed a unique domain inversion in $Co_2SiO_4$ by applying a magnetic field, which cannot be identified by macroscopic property measurements. This response can be explained by a trilinear coupling mediated by the piezomagnetic effect. This suggests that $E_T$ domains can be controlled through the simultaneous application of shear stress and a magnetic field. Furthermore, in systems where the piezomagnetic effect is permitted, magnetic linear dichroism emerges [53], and as its inverse effect, $E_T$ domains may be controlled via irradiation with linearly polarized light in a magnetic field [54].

**Acknowledgments**

We thank K. Kimura and T. Aoyama for fruitful discussions. The images of crystal structures were drawn using the software VESTA [55]. This work was supported by JSPS 271 KAKENHI Grants No. JP19H05823 and No. JP21H04436.

## APPENDIX A: METHODS

### 1. Sample preparation

A $Co_2SiO_4$ single crystal was grown by the floating zone method [56]. The obtained crystal was oriented using Laue x-ray diffraction, and three plate-shaped samples with the widest faces perpendicular to the $a$, $b$, and $c$ axes were prepared. For the transmittance optical measurements under an electric field, the plate-shaped samples were polished to a thickness of 50–90 μm, and indium–tin oxide was sputtered onto the widest faces of the samples to form a pair of transparent electrodes. Magnetization was measured using a commercial physical property measurement system (PPMS, Quantum Design).

### 2. Spatial distribution measurements of *E*-induced NDD

Two-dimensional maps of the $E$-induced NDD were obtained by using an electric-field-modulation imaging technique [45]. In these measurements, a square-wave voltage was applied to a sample, and transmission microscope images were captured using an sCMOS camera (pco edge 5.5, PCO) under alternating positive ($+V$) and negative ($-V$) voltages. Then, the difference in signals under $+V$ and $-V$ [$\Delta I = I(+V) - I(-V)$] divided by their average ($I$) was calculated for each pixel. Using equations (1) and (2), $\Delta I/I$ may be expressed as

$$\frac{\Delta I}{I} \cong \frac{I_0 e^{-\alpha_0 d}(1-\Delta\alpha d) - I_0 e^{-\alpha_0 d}(1+\Delta\alpha d)}{\{I_0 e^{-\alpha_0 d}(1+\Delta\alpha d) + I_0 e^{-\alpha_0 d}(1-\Delta\alpha d)\}/2} = -2\Delta\alpha d = -2\beta V \quad (3)$$

To obtain spatial distributions of small signals of $\Delta I/I$ while minimizing noise, large numbers (5,000 ~ 15,000) of $\Delta I/I$ maps were captured and averaged. A square-wave bias voltage was applied at a frequency of 20 Hz, and images were captured at 40 fps.

The sample temperature was controlled using a liquid He flow cryostat (Microstat He, Oxford Instruments). Domain imaging under a magnetic field was performed using an electromagnet (3470, GMW associate).

### 3. Spectral measurements of optical absorption and *E*-induced NDD

Optical absorption spectra and *E*-induced NDD spectra were measured using a supercontinuum laser (SC-Pro, YSL Photonics) and an acousto-optic wavelength tunable filter (AOTF-PRO, YSL Photonics). The spectra were measured by varying the wavelength of the incident linearly polarized light in 1 nm increments. In the absorption spectrum measurements, the intensity of light transmitted through the sample was measured and normalized using the intensity of light transmitted through only the window. In the *E*-induced NDD spectrum measurements, signals were detected using a lock-in technique. A sinusoidal voltage $V_0\sin(2\pi ft)$ was applied, where $V_0 = 150$ V and $f = 999$ Hz. Under such a sinusoidal voltage, the intensity of the transmitted light may be expressed as

$$I_0 e^{-(\alpha_0+\beta V_0 \sin(2\pi f t)/d)d} \approx I_0 e^{-\alpha_0 d}\left(1-\beta V_0 \sin(2\pi f t)\right) \quad (4).$$

Then, the intensity of transmitted light oscillating at the frequency $f$ was detected by using a lock-in amplifier, and $\Delta\alpha \equiv \beta V_0/d$ was calculated. The *E*-induced NDD spectra were measured at both $E_T$+ and $E_T$– domains by focusing the laser to a point smaller than the domain size (beam size ~ 50 μm). In the same manner with the spatial distribution measurements, the sample temperature was controlled using the liquid He flow cryostat.

## APPENDIX B: CALCULATION OF *E*-INDUCED NDD FROM THE DOMAIN IMAGES

The white boxes in Figs. 8(a), 8(b), and 8(c) show the areas where the average of $\Delta\alpha d$ was obtained to calculate the voltage dependence of *E*-induced NDD in Figs. 4(d), 4(h), and 4(l) in the main text, respectively.

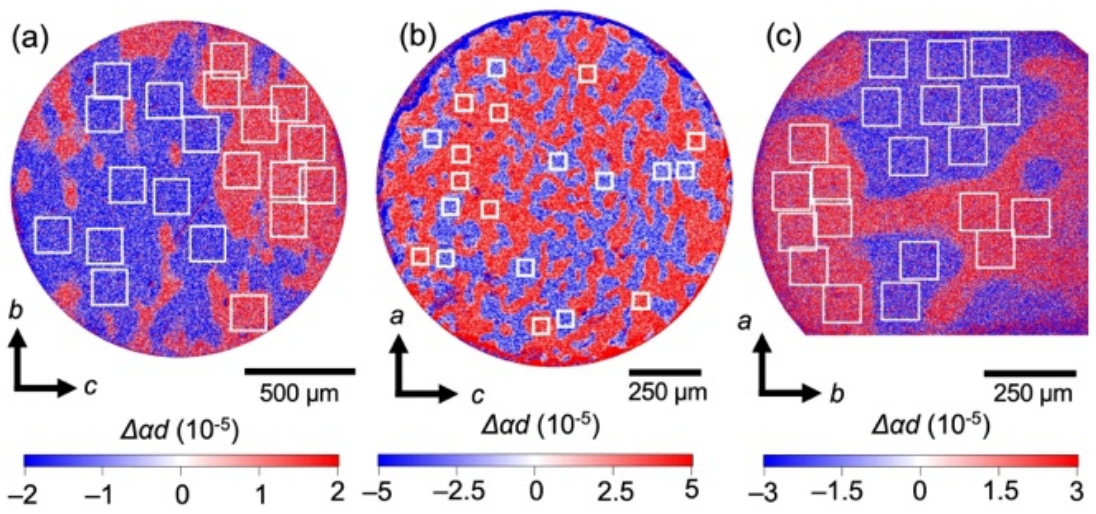

FIG. 8. Sampling positions for calculating the voltage dependence of *E*-induced NDD.

## APPENDIX C: TEMPERATURE DEPENDENCE OF *E*-INDUCED NDD

The temperature dependence of $\Delta\alpha$ at 150 V was obtained by focusing a laser beam on a single domain region and using a lock-in technique, with the sample temperature increasing at a rate of 1 K/min (Fig. 9). Considering the results of the spectral measurements (Fig. 10 in Appendix D), the wavelengths that maximized $\Delta\alpha$ were selected [588 nm for Figs. 9(a) and 9(f), 587 nm for Figs. 9(b) and 9(e), 589 nm for Figs. 9(c) and 9(d)]. Except for Fig. 9(a), the signals of *E*-induced NDD disappear around 50 K, corresponding to $T_N$. In Fig. 9(a), the signal disappears at a lower temperature (≈ 47 K), which is attributed to heating caused by irradiation with the focused laser.

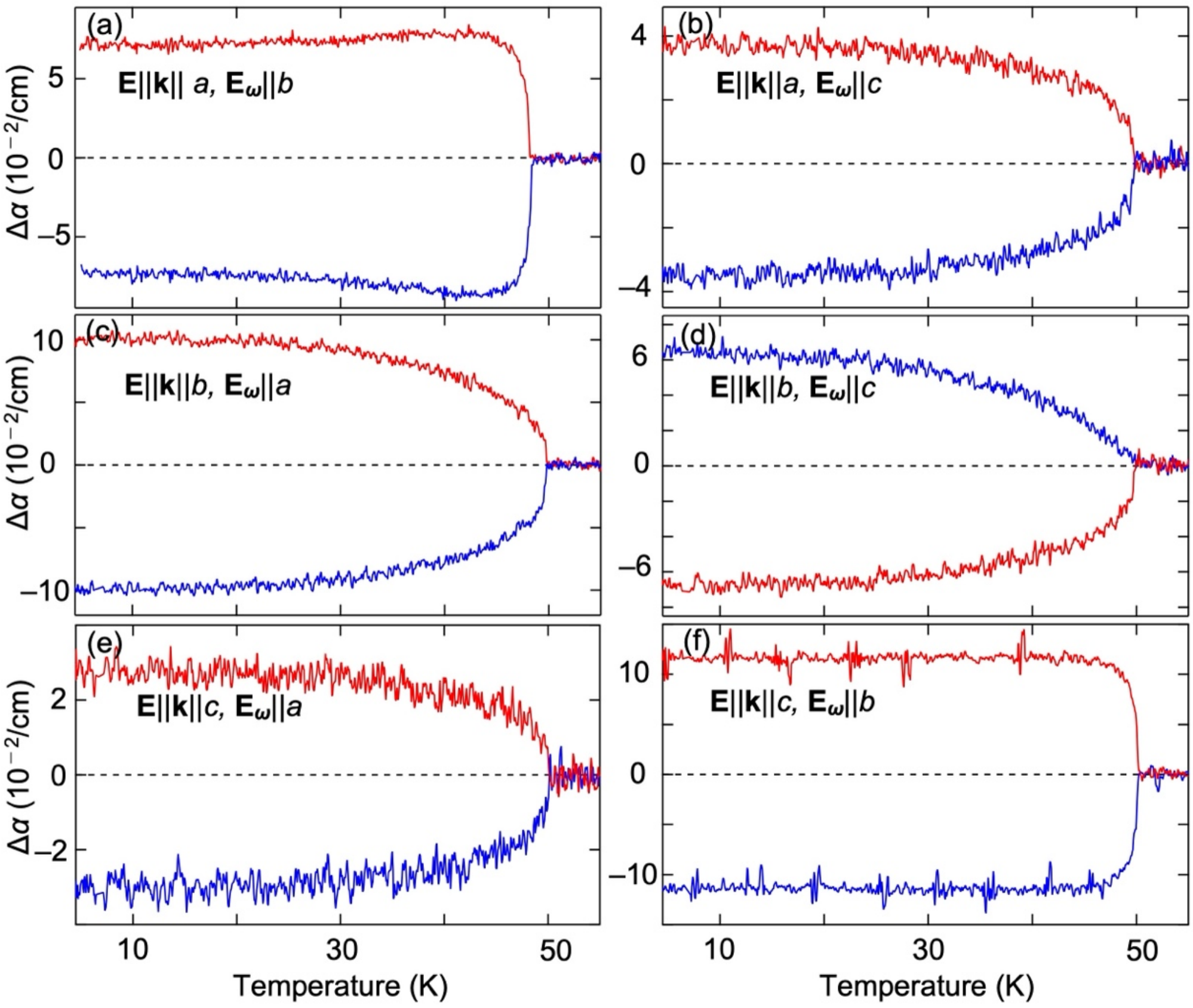

FIG. 9. Temperature dependence of $E$-induced NDD. The temperature dependence of $\Delta\alpha$ at 150 V was obtained by focusing a laser beam on a single domain region and using a lock-in technique, with the sample temperature increasing at a rate of 1 K/min. Following the results of the spectral measurements (Fig. 10 in Appendix D), the wavelengths that maximized $\Delta\alpha$ were selected [588 nm for (a),(f), 587 nm for (b),(e), 589 nm for (c),(d)]. Panels (a) [(b)] show the temperature dependence of $\alpha$ and $\Delta\alpha$ for $b$[$c$]-polarized light ($\mathbf{E}_\omega||b[c]$) propagating along the $a$ axis ($\mathbf{k}||a$). In the same manner, panels (c)[(d)] show the temperature dependence for $\mathbf{k}||b$ and $\mathbf{E}_\omega||a[c]$, while panels (e)[(f)] show those for $\mathbf{k}||c$ and $\mathbf{E}_\omega||a[b]$. The red and blue correspond to the data obtained with either $E_T$+ or $E_T$– domains. The results in two different $\mathbf{E}_\omega$ for the same $\mathbf{k}$ obtained at the same domain state are represented by the same color.

## APPENDIX D: ENERGY DEPENDENCE OF *E*-INDUCED NDD

Here we discuss the details of $E$-induced NDD spectrum. The obtained $E$-induced NDD spectra are shown in Figs. 10(d)-10(f) and 10(j)-10(l) along with the absorption spectra [Figs. 10(a)-10(c) and 10(g)-10(i)]. Here, the measurements in the two polarization settings of each **k** direction were performed for the same domain state, and the red and blue dots denote the data obtained in either $E_T$+ or $E_T$– domains. Figures10(a) and 10(d) [10(g) and 10(j)] show the spectra of $\alpha$ and $\Delta\alpha$ for $b$[$c$]-polarized light ($\mathbf{E}_\omega||b[c]$) propagating along the $a$ axis ($\mathbf{k}||a$), respectively. In the same manner, Figs.10(b) and 10(e) [10(h) and 10(k)] show the spectra for $\mathbf{k}||c$ and $\mathbf{E}_\omega||b[a]$, while Figs.10(c) and 10(f) [10(i) and 10(l)] show those for $\mathbf{k}||b$ and $\mathbf{E}_\omega||a[c]$. The $E$-induced NDD spectra taken at the opposite domains show a complete sign reversal [compare red and blue lines in Figs. 10(d)-10(f) and Figs. 10(j)-10(l)]. In $Co_2SiO_4$, the absorption in the visible light region is mainly due to the $Co^{2+}$ $d$-$d$ transition [46]. Among the two Co sites, the Co1 site in a centrosymmetric field does not significantly contribute to the absorption [46]. Thus, we focus on the Co2 site in the following discussion. The absorption peaks observed around 2.0–2.6 eV are ascribed to the $d$-$d$ transition from the ground state $T_{1g}$ to the excited $T_{1g}$ state. Here, $T_{1g}$ refers to the state split by the cubic crystal field. Because of the effect of the distorted crystal field, it is further split into roughly three states. Around the same energy, the $E$-induced NDD shows complicated peak structures. In the following discussion, we do not delve into the details of all the peaks but consider a model to explain the origin of $E$-induced NDD in each setting from the overall features of the spectra.

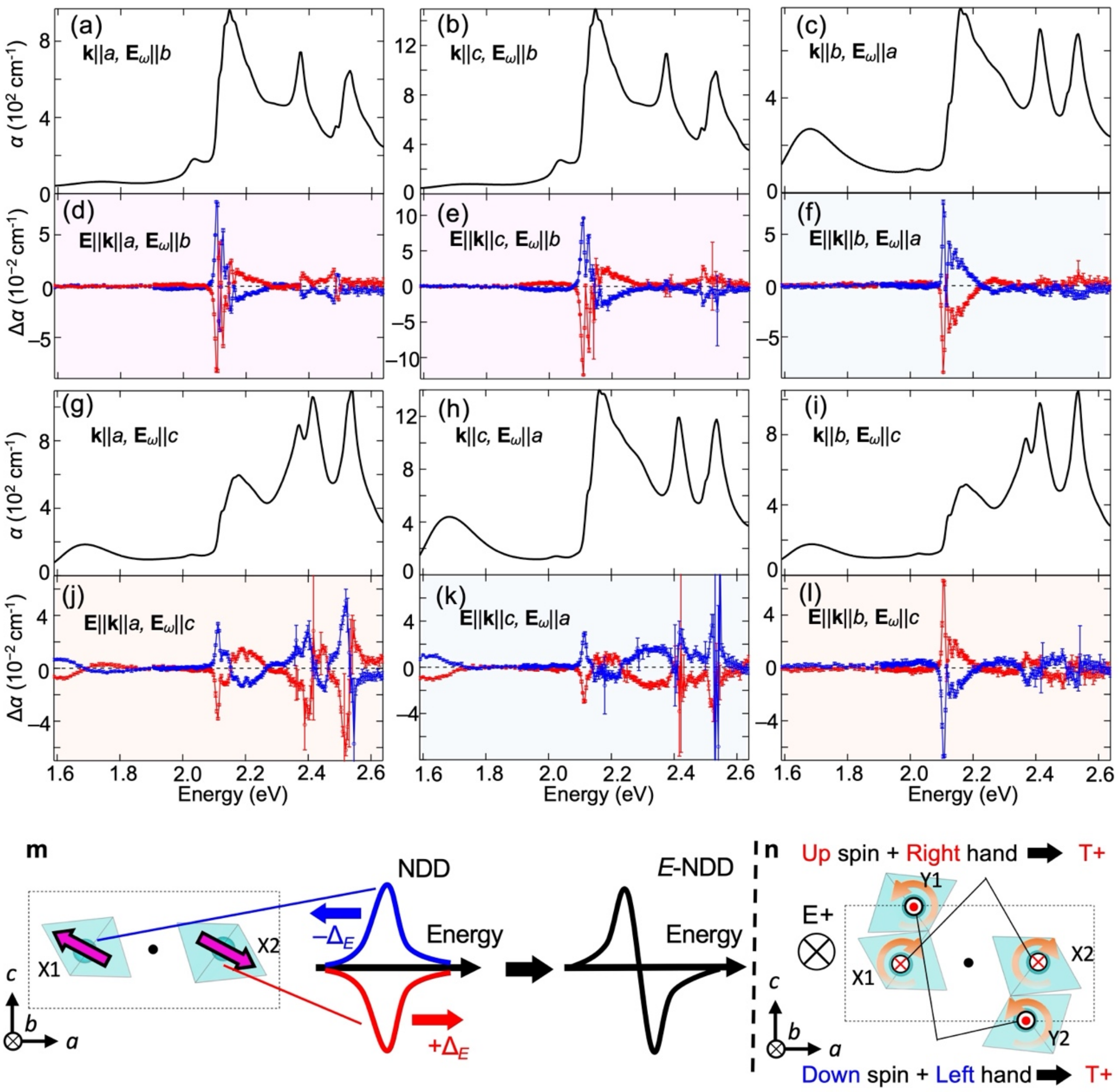

FIG. 10. Spectra of $E$-induced NDD**.** (a)-(l) Energy dependence of an absorption coefficient $\alpha$ [(a)-(c) and (g)-(i)] and $E$-induced NDD $\Delta\alpha$ under 150 V [(d)-(f) and (j)-(l)] at 4 K. Here, for comparison with $\alpha$, $E$-induced NDD is shown as $\Delta\alpha$, not $\Delta\alpha d$. The red and blue dots in the $E$-induced NDD spectra correspond to the data obtained for either $E_T+$ or $E_T-$ domains. The results in two different $\mathbf{E}_\omega$ for the same $\mathbf{k}$ obtained at the same domain state are represented by the same color. The $\Delta\alpha$ spectra for the same direction of $\mathbf{E}_\omega$ are highlighted in the same color. Error bars show standard deviations. (m) Model for the $E$-induced NDD in the case of $\mathbf{E}||\mathbf{k}||a$ or $c$. Although the local toroidal moments at X sites (magenta arrows) locally induce NDD, the contributions of X1 and X2 sites cancel each other out in the absence of an electric field. However, under an electric field, the resonance energy of the $E$-induced NDD is shifted in opposite sign ($\pm\Delta_E$), which induces finite NDD. (n) Model for the $E$-induced toroidal moment in the case of $\mathbf{E}||\mathbf{k}||b$. An electric field induces chirality at sites X and Y with opposite handedness. The toroidal moment is induced by the coupling of local spins and chirality. As the spins of sites X and Y are opposite, the signs of the toroidal moments at these sites are the same.

An important feature is that the spectra at $\mathbf{E}||\mathbf{k}||a$ and $\mathbf{E}_\omega||b$ [Fig. 10(d)] and those at $\mathbf{E}||\mathbf{k}||c$ and $\mathbf{E}_\omega||b$ [Fig. 10(e)] are quite similar, whereas those at $\mathbf{E}||\mathbf{k}||a$ and $\mathbf{E}_\omega||c$ [Fig. 10(j)] significantly differ from those at $\mathbf{E}||\mathbf{k}||b$ and $\mathbf{E}_\omega||c$ [Fig. 10(l)], and those at $\mathbf{E}||\mathbf{k}||c$ and $\mathbf{E}_\omega||a$ [Fig. 10(k)] from those at $\mathbf{E}||\mathbf{k}||b$ and $\mathbf{E}_\omega||a$ [Fig. 10(f)]. The $E$-induced NDD spectra at $\mathbf{E}||\mathbf{k}||a,c$ have finer peak structures, where positive and negative signal swings are sharper than those at $\mathbf{E}||\mathbf{k}||b$. Because no significant differences are observed in the absorption spectra [compare Fig. 10(g) with Fig. 10(i)], the magnetic structure rather than structural anisotropy should determine the difference in the $E$-induced NDD spectra. This can be considered in a model based on the local magnetic toroidal moments as follows.

Let us first discuss the model for the $E$-induced NDD spectra at $\mathbf{E}||\mathbf{k}||b$. As illustrated in Fig. 10(n), $Co^{2+}$ spins at the Co2 site are parallel to the $b$ axis so that a toroidal moment parallel to the $b$ axis seems not to be induced even if **E** is applied. At finite **E**, however, the mirror plane perpendicular to the $b$ axis at each site is broken, resulting in a chiral state. When the system is chiral, a toroidal moment parallel to a magnetic moment is induced because of magneto-chiral coupling [57,58]. Now, the two sites (X1 and X2) with up spins have the same chirality (let us assume a right-handed structure) when **E** is applied, and the other two sites (Y1 and Y2) with down spins have left-handed chirality. The X sites have up spin and right-handed chirality, and the Y sites have down spin and left-handed chirality, i.e., the X and Y sites exhibit opposite spins and chirality. Because a local magnetic toroidal moment **t** is the product of spin and chirality, the sites show the same polarity of induced **t**. Accordingly, the appearance of $E$-induced NDD at $\mathbf{E}||\mathbf{k}||b$ can be interpreted as the sum of **t** becoming finite under **E**. With such a finite toroidal moment, the photon energy ($\hbar\omega$) dependence of NDD around a single excitation energy $\hbar\omega_{n0}$ is described as [59,60]

$$\Delta\alpha(\omega) \propto \frac{\mathrm{Re}\left[<0\left|m_\beta\right|n><n|p_\alpha|0>\right]}{(\omega-\omega_{n0})^2+\delta_n^2}, \tag{5}$$

where $<0\left|m_\beta\right|n>$ and $<n|p_\alpha|0>$ are matrix elements of magnetic ($m_\beta$) and electric ($p_\alpha$) dipole operators, respectively, taken between the ground $|0>$ and the excited states $|n>$ separated by $\hbar\omega_{n0}$ energy; $\alpha$ and $\beta$ specify the directions of the electric and magnetic fields of light, respectively; $\delta_n$ is the inverse lifetime of the excited state $|n>$. According to this equation, $\Delta\alpha(\omega)$ has an absorptive structure with a single peak at $\omega_{n0}$, which is roughly consistent with the spectra shown in Fig. 10(f) and 10(l).

Let us now consider $E$-induced NDD at $\mathbf{E}||\mathbf{k}||a,c$. As mentioned above, **t** at each Co2 site is finite along the $a$ and $c$ axes. Although the sum of **t** in a unit cell is canceled out, the application of **E** along $a$ or $c$ induces a net toroidal moment along the same direction of **E**. However, this net toroidal moment is not a major contributor to the NDD at the resonance energy. Instead, the difference between mutually antiparallel local toroidal moments is a determining factor. Let us take the X site as an example. Owing to the presence of local toroidal moments, NDD occurs locally, but the effects cancel each other out at X1 and X2. However, when **E** is applied, the excitation energy that induces $\Delta\alpha$ changes linearly with respect to **E** ($\Delta_E \propto |\mathbf{E}|$), where the sign of $\Delta_E$ is opposite at the X1 and X2 sites (pseudo-Stark effect) [61,62]. As a result, the peak energies of NDD at the X1 and X2 sites become slightly different, so that they no longer cancel out each other, resulting in a finite NDD. The same process occurs at the Y1 and Y2 sites. This behavior, taking the pseudo-Stark effect into account, may be described as follows:

$$\begin{aligned}\Delta\alpha(\omega) &= \Delta\alpha^{\mathrm{X1}}(\omega)-\Delta\alpha^{\mathrm{X2}}(\omega)\\ &\propto \mathrm{Re}\left[<0\left|m_\beta\right|n><n|p_\alpha|0>\right]\left\{\frac{1}{[\omega-(\omega_{n0}+\Delta_E)]^2+\delta_n^2}-\frac{1}{[\omega-(\omega_{n0}-\Delta_E)]^2+\delta_n^2}\right\}\\ &\approx \mathrm{Re}\left[<0\left|m_\beta\right|n><n|p_\alpha|0>\right]\frac{\omega-\omega_{n0}}{[(\omega-\omega_{n0})^2+\delta^2]^2}\Delta_E. \end{aligned} \tag{6}$$

In this case, $\Delta\alpha(\omega)$ is proportional to $1/(\omega-\omega_{n0})^3$ and shows a dispersive structure with two peaks of opposite signs around the resonance frequency $\omega_{n0}$, which is roughly consistent with the spectra shown in Figs. 10(d), 10(e), 10(j), and 10(k). Therefore, in the resonance energy region, the existence of these antiparallel toroidal moments has a stronger effect on $E$-induced NDD than the finite sum of toroidal moments under **E**.

Considering the above, the difference in the shape of the $E$-induced NDD spectra at $\mathbf{E}||\mathbf{k}||a,c$ and $\mathbf{E}||\mathbf{k}||b$ can be roughly explained. However, this explanation does not take into account the contribution of the spins on the Co1 site and the effect of orbital hybridization due to spin-orbit coupling [63], which can be another factor in the dispersive peak structure, as a more detailed spectral analysis would be required to account for them.

### APPENDIX E: $E_T$ domain inversion by sweeping a magnetic field

In the (100) plane at 49 K (just below $T_N$), we performed the domain imaging with sweeping the magnetic field along the *c* axis as 0 T → 200 mT →−200 mT→200 mT. Figure 11 shows the summary of the results. By applying just 50 mT, the domain patterns changed, as the number of domain walls perpendicular to the *c* axis (dw⊥*c*) increased [compare Fig. 11(a) with 11(b)]. When the magnetic field was increased to 200 mT, this tendency became more pronounced [Fig. 11(c)]. One finds that the domain contrast at 200 mT is somewhat whitish, which suggests that tiny regions with weak signals, for example, tiny multi-$E_T$ domains smaller than the microscope resolution (several μm), spread throughout the sample. With decreasing *H*, the domain structure started changing again in between 50 mT and 0 T. After flipping the direction of the magnetic field, the same tendency of the increasing dw⊥*c* with increasing *H* was observed. However, the domain contrast gets inverted with maintaining the pattern from that under +*H* [compare Fig. 11(e) with Fig. 11(c)]. The *H* dependence of $\Delta\alpha d$ averaged in one selected area surrounded by a black box in Fig. 11(a) shows a hysteresis loop as shown in Fig. 11(f). The coercive magnetic field is 10 ~ 20 mT, which does not contradict the magnetization measurement where the coercive magnetic field is 5 ~ 10 mT at 49 K [see Fig. 11(g)]. The magnitude of $|M|$ increases in proportion to $|H|$, whereas the magnitude of $|\Delta\alpha d|$ begins to decrease above $|\mu_0 H|$ = 20 mT. This is most likely because as *H* is increased, regions with smaller strains respond to *H*, forming multidomains that are smaller than the resolution limit of the microscope (for the relationship between strain and magnetic field, see Main text).

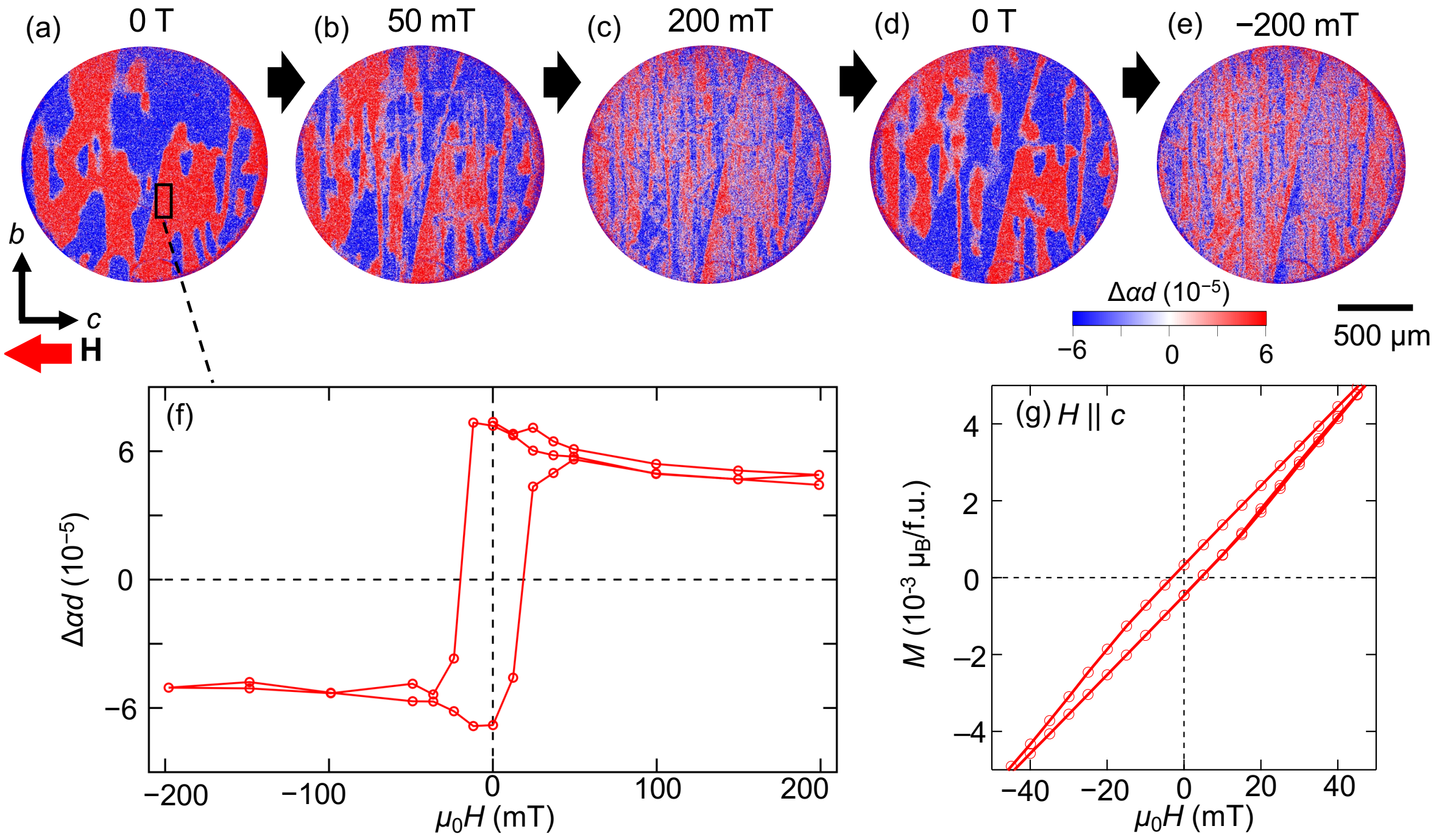

FIG. 11. $E_T$ domain inversion by sweeping a magnetic field. (a)-(e) $E_T$ domains in the (100) plane obtained while sweeping the magnetic field *H* applied along the *c* axis as 0 T→200 mT→−200 mT→200 mT. The measurements were performed at 49 K (< $T_N$). The magnetic field was kept constant during each domain imaging. The *b*-polarized light with the wavelength of 590 nm was used for the imaging and the applied voltage was 200 V. When one compares the domains at 200 mT [(c)] and −200 mT [(e)], the pattern is the same, but the contrast is inverted. (f) *H* dependence of $\Delta\alpha d$ averaged in the areas surrounded by the black box in panel (a). It shows a hysteresis behavior. (g) Magnetization curves at 49 K for *H* along the *c* axis. See also Appendix. F.

## APPENDIX F: DETAILED MAGNETIZATION MEASUREMENTS

Temperature dependence of spontaneous magnetization [Figs. 12(a)-12(c)] and magnetization curve at 49 K (just below $T_N$) [Figs. 12(d)-12(f)] were measured along the *a*, *b,* and *c* axes.

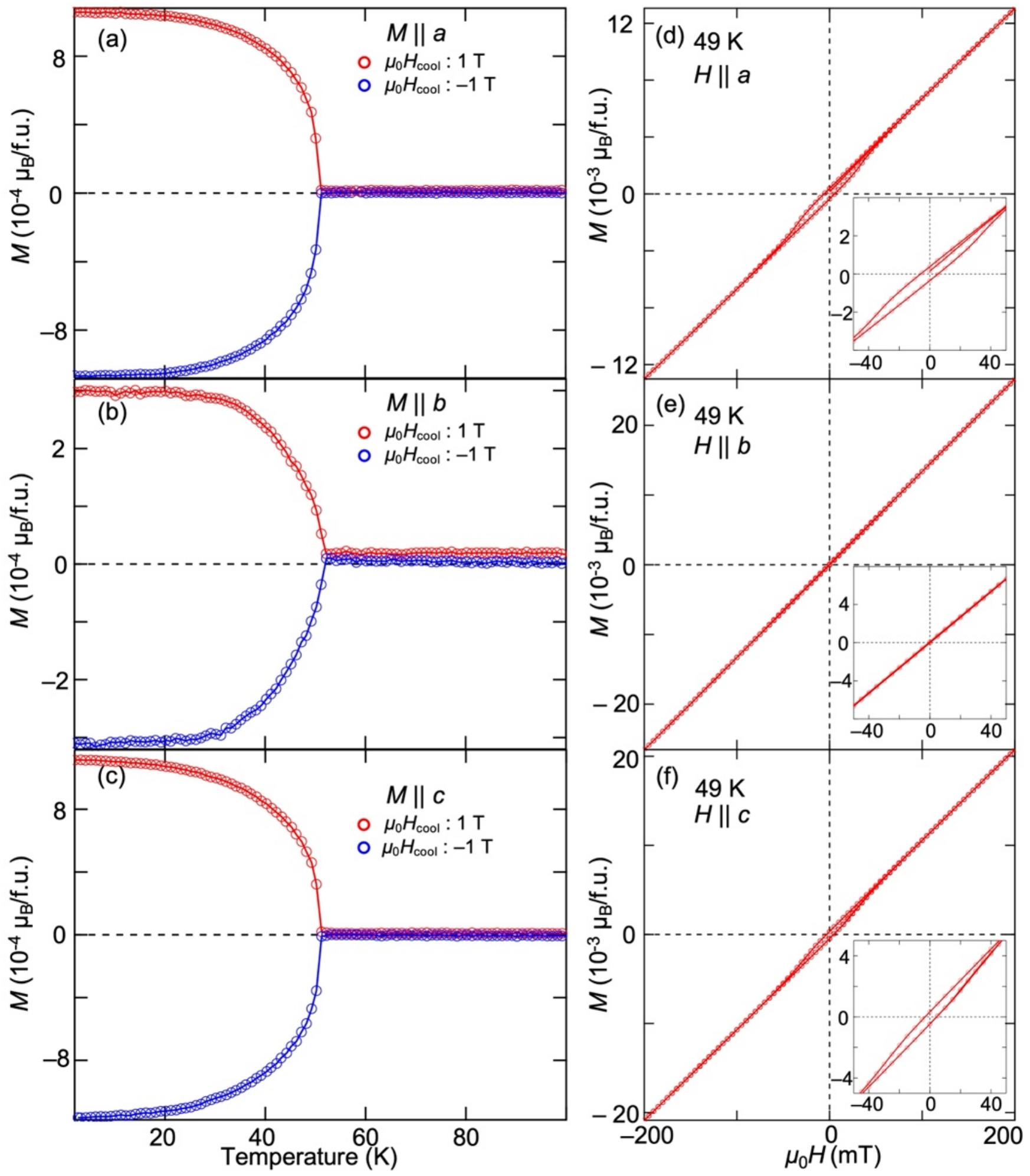

FIG. 12. (a)-(c) Temperature dependence of spontaneous magnetization ($M$) along the $a$[(a)], $b$[(b)], and $c$[(c)] axes. The red and blue dots show the results measured after cooling the sample across $T_N$ in a magnetic field ($H$) of 1 and –1 T, respectively. A small magnetic field up to ±0.001 T was applied during the measurements to suppress residual magnetic field. The offsets observed above $T_N$ were attributed to residual magnetic field. (d)-(f) Magnetization curves at 49 K (just below $T_N$) for $H$ along the $a$[(d)], $b$[(e)], and $c$[(f)] axes. The insets show enlarged views of the range from –50 to 50 mT. Small spontaneous magnetization and hysteresis behaviors are observed for $H$ along the $a$ and $c$ axes, but not for $H$ along the $b$ axis.

[1] L. Néel, *Magnetism and Local Molecular Field*, Science **174**, 985 (1971).

[2] I. E. Dzyaloshinskii, *On the Magneto-Electrical Effect in Antiferromagnetics*, Zh. Exp. Teor. Fiz. **33**, 881 (1959).

[3] V. J. Folen, G. T. Rado, and E. W. Dtalder, *Anisotropy of the Mangetoelectric Effect in $Cr_2O_3$*, Phys. Rev. Lett. **6**, 607 (1961).

[4] A. N. Spaldin and M. Fiebig, *The Renaissance of Magnetoelectric Multiferroics*, Science **309**, 391 (2005).

[5] Y. Tokura and N. Nagaosa, *Nonreciprocal Responses from Non-Centrosymmetric Quantum Materials*, Nat. Commun. **9**, 3740 (2018).

[6] S. Nakatsuji, N. Kiyohara, and T. Higo, *Large Anomalous Hall Effect in a Non-Collinear Antiferromagnet at Room Temperature*, Nature **527**, 212 (2015).

[7] W. Feng, G. Y. Guo, J. Zhou, Y. Yao, and Q. Niu, *Large Magneto-Optical Kerr Effect in Noncollinear Antiferromagnets $Mn_3X$ (X=Rh, Ir, Pt)*, Phys. Rev. B **92**, 144426 (2015).

[8] T. Higo et al., *Large Magneto-Optical Kerr Effect and Imaging of Magnetic Octupole Domains in an Antiferromagnetic Metal*, Nat. Photonics **12**, 73 (2018).

[9] M. Naka, S. Hayami, H. Kusunose, Y. Yanagi, Y. Motome, and H. Seo, *Spin Current Generation in Organic Antiferromagnets*, Nat. Commun. **10**, 4305 (2019).

[10] L. Šmejkal, R. González-Hernández, T. Jungwirth, and J. Sinova, *Crystal Time-Reversal Symmetry Breaking and Spontaneous Hall Effect in Collinear Antiferromagnets*, Sci. Adv. **6**, eaaz8809 (2020).

[11] Z. Feng et al., *An Anomalous Hall Effect in Altermagnetic Ruthenium Dioxide*, Nature Electronics **5**, 735 (2022).

[12] L. Šmejkal, A. H. MacDonald, J. Sinova, S. Nakatsuji, and T. Jungwirth, *Anomalous Hall Antiferromagnets*, Nature Reviews Materials **7**, 482 (2022).

[13] J. Krempaský et al., *Altermagnetic Lifting of Kramers Spin Degeneracy*, Nature **626**, 517 (2024).

[14] L. Šmejkal, J. Sinova, and T. Jungwirth, *Beyond Conventional Ferromagnetism and Antiferromagnetism: A Phase with Nonrelativistic Spin and Crystal Rotation Symmetry*, Phys. Rev. X **12**, 031042 (2022).

[15] L. Šmejkal, J. Sinova, and T. Jungwirth, *Emerging Research Landscape of Altermagnetism*, Phys. Rev. X **12**, 040501 (2022).

[16] H. Schmid, *On Ferrotoroidics and Electrotoroidic, Magnetotoroidic and Piezotoroidic Effects*, Ferroelectrics **252**, 41 (2001).

[17] H. Schmid, *Some Symmetry Aspects of Ferroics and Single Phase Multiferroics*, J. Phys. Condens. Matter **20**, 434201 (2008).

[18] V. L. Ginzburg, A. A. Gorbatsevich, Y. V. Kopayev, and B. A. Volkov, *On the Problem of Superdiamagnetism*, Solid State Commun. **50**, 339 (1984).

[19] V. M. Dubovik and V. V. Tugushev, *Toroid Moments in Electrodynamics and Solid-State Physics*, Phys. Rep. **187**, 145 (1990).

[20] A. A. Gorbatsevich and Y. V. Kopaev, *Toroidal Order in Crystals*, Ferroelectrics **161**, 321 (1994).

[21] B. B. Van Aken, J. P. Rivera, H. Schmid, and M. Fiebig, *Observation of Ferrotoroidic Domains*, Nature **449**, 702 (2007).

[22] N. A. Spaldin, M. Fiebig, and M. Mostovoy, *The Toroidal Moment in Condensed-Matter Physics and Its Relation to the Magnetoelectric Effect*, J. Phys. Condens. Matter **20**, 434203 (2008).

[23] G. L. J. A. Rikken, C. Strohm, and P. Wyder, *Observation of Magnetoelectric Directional Anisotropy*, Phys. Rev. Lett. **89**, 133005 (2002).

[24] S. Hayami and H. Kusunose, *Time-Reversal Switching Responses in Antiferromagnets*, Phys. Rev. B **108**, L140409 (2023).

[25] X. Xu, F.-T. Huang, and S.-W. Cheong, *Magnetic Toroidicity*, J. Phys. Condens. Matter **36**, 203002 (2024).

[26] M. Yatsushiro, H. Kusunose, and S. Hayami, *Multipole Classification in 122 Magnetic Point Groups for Unified Understanding of Multiferroic Responses and Transport Phenomena*, Phys. Rev. B **104**, 054412 (2021).

[27] R. Winkler and U. Zülicke, *Theory of Electric, Magnetic, and Toroidal Polarizations in Crystalline Solids with Applications to Hexagonal Lonsdaleite and Cubic Diamond*, Phys. Rev. B **107**, 155201 (2023).

[28] S. Bhowal and N. A. Spaldin, *Ferroically Ordered Magnetic Octupoles in d-Wave Altermagnets*, Phys. Rev. X **14**, 011019 (2024).

[29] J. H. Jung, M. Matsubara, T. Arima, J. P. He, Y. Kaneko, and Y. Tokura, *Optical Magnetoelectric Effect in the Polar $GaFeO_3$ Ferrimagnet*, Phys. Rev. Lett. **93**, 037403 (2004).

[30] M. Kubota, T. Arima, Y. Kaneko, J. P. He, X. Z. Yu, and Y. Tokura, *X-Ray Directional*

*Dichroism of a Polar Ferrimagnet*, Phys. Rev. Lett. **92**, 137401 (2004).

[31] S. Reschke et al., *Confirming the Trilinear Form of the Optical Magnetoelectric Effect in the Polar Honeycomb Antiferromagnet $Co_2Mo_3O_8$*, npj Quantum Mater. **7**, 1 (2022).

[32] K. Kimura and T. Kimura, *Nonvolatile Switching of Large Nonreciprocal Optical Absorption at Shortwave Infrared Wavelengths*, Phys. Rev. Lett. **132**, 036901 (2024).

[33] T. Sato, N. Abe, Y. Tokunaga, and T. H. Arima, *Antiferromagnetic Domain Wall Dynamics in Magnetoelectric $MnTiO_3$ Studied by Optical Imaging*, Phys. Rev. B **105**, 094417 (2022).

[34] K. Kimura, Y. Otake, and T. Kimura, *Visualizing Rotation and Reversal of the Néel Vector through Antiferromagnetic Trichroism*, Nat. Commun. **13**, 697 (2022).

[35] S. Ghose and C. Wan, *Strong Site Preference of $Co^{2+}$ in Olivine, $Co_{1.10}Mg_{0.90}SiO_4$*, Contrib. Mineral. Petrol. **47**, 131 (1974).

[36] N. Morimoto, M. Tokonami, M. Watanabe, and K. Koto, *Crystal Structures of Three Polymorphs of $Co_2SiO_4$*, Am. Mineral. **59**, 475 (1974).

[37] A. Sazonov, V. Hutanu, M. Meven, G. Heger, T. Hansen, and A. Senyshyn, *Anomalous Thermal Expansion of Cobalt Olivine, $Co_2SiO_4$, at Low Temperatures*, J. Appl. Crystallogr. **43**, 720 (2010).

[38] W. Lottermoser and H. Fuess, *Magnetic structure of the orthosilicates $Mn_2SiO_4$ and $Co_2SiO_4$*, phys. stat. sol. (a) **109**, 589 (1988).

[39] O. Ballet, H. Fuess, K. Wacker, E. Untersteller, W. Treutmann, E. Hellner, and S. Hosoya, *Magnetisation Measurements of the Synthetic Olivine Single Crystals $A_2SiO_4$ with A=Mn, Fe or Co*, J. Phys. Condens. Matter **1**, 4955 (1989).

[40] A. Sazonov, M. Meven, V. Hutanu, G. Heger, T. Hansen, and A. Gukasov, *Magnetic Behaviour of Synthetic $Co_2SiO_4$*, Acta Crystallogr. B **65**, 664 (2009).

[41] R. R. Birss, *Symmetry & Magnetism* (North-holland publishing company, Amsterdam, 1966).

[42] A. Smolyanyuk, L. Šmejkal, and I. I. Mazin, *A Tool to Check Whether a Symmetry-Compensated Collinear Magnetic Material Is Antiferro- or Altermagnetic*, SciPost Phys. Codebases, 30 (2024).

[43] S.-W. Cheong and F.-T. Huang, *Altermagnetism with Non-Collinear Spins*, npj Quantum Materials **9**, 13 (2024).

[44] T. Hayashida, K. Arakawa, T. Oshima, K. Kimura, and T. Kimura, *Observation of Antiferromagnetic Domains in $Cr_2O_3$ Using Nonreciprocal Optical Effects*, Phys. Rev. Res. **4**, 043063 (2022).

[45] Y. Uemura, S. Arai, J. Tsutsumi, S. Matsuoka, H. Yamada, R. Kumai, S. Horiuchi, A. Sawa, and T. Hasegawa, *Field-Modulation Imaging of Ferroelectric Domains in Molecular Single-Crystal Films*, Phys. Rev. Appl. **11**, 014046 (2019).

[46] M. N. Taran and G. R. Rossman, *Optical Spectra of $Co^{2+}$ in Three Synthetic Silicate Minerals*, Am. Mineral. **86**, 889 (2001).

[47] T. Honda, J. S. White, A. B. Harris, L. C. Chapon, A. Fennell, B. Roessli, O. Zaharko, Y. Murakami, M. Kenzelmann, and T. Kimura, *Coupled Multiferroic Domain Switching in the Canted Conical Spin Spiral System $Mn_2GeO_4$*, Nat. Commun. **8**, 15457 (2017).

[48] N. Leo et al., *Magnetoelectric Inversion of Domain Patterns*, Nature **560**, 466 (2018).

[49] E. Hassanpour, Y. Zemp, Y. Tokunaga, Y. Taguchi, Y. Tokura, T. Lottermoser, M. Fiebig, and M. C. Weber, *Magnetoelectric Transfer of a Domain Pattern*, Science **377**, 1109 (2022).

[50] T. Nakajima, Y. Tokunaga, Y. Taguchi, Y. Tokura, and T.-H. Arima, *Piezomagnetoelectric Effect of Spin Origin in Dysprosium Orthoferrite*, Phys. Rev. Lett. **115**, 197205 (2015).

[51] I. E. Dzialoshinskii, *The Problem of Piezomagnetism*, Sov. Phys. JETP **6**, 621 (1958).

[52] A. S. Borovik-Romanov, *Piezomagnetism in Antiferromagnetic Cobalt and Manganese Fluorides*, J. Exptl. Theoret. Phys. (U.S.S.R.) **36**, 1088 (1959).

[53] A. S. Borovik-romanov, *Piezomagnetism, Linear Magnetostriction and Magnetooptic Effect*, Ferroelectrics **162**, 153 (1994).

[54] T. Higuchi and M. Kuwata-Gonokami, *Control of Antiferromagnetic Domain Distribution via Polarization-Dependent Optical Annealing*, Nat. Commun. **7**, 10720 (2016).

[55] K. Momma and F. Izumi, *Vesta3 for Threedimensional Visualization of Crysal, Volumetric and Morphology Data*, J. Appl. Crystallogr. **44**, 1272 (2011).

[56] Q. Tang and R. Dieckmann, *Floating-Zone Growth and Characterization of Single Crystals of Cobalt Orthosilicate, $Co_2SiO_4$*, J. Cryst. Growth **317**, 119 (2011).

[57] T. Sato, N. Abe, S. Kimura, Y. Tokunaga, and T.-H. Arima, *Magnetochiral Dichroism in a Collinear Antiferromagnet with No Magnetization*, Phys. Rev. Lett. **124**, 217402 (2020).

[58] S.-W. Cheong, S. Lim, K. Du, and F.-T. Huang, *Permutable SOS ( Symmetry Operational Similarity )*, npj Quantum Mater. **6**, 58 (2021).
[59] I. Kézsmárki et al., *One-Way Transparency of Four-Coloured Spin-Wave Excitations in Multiferroic Materials*, Nat. Commun. **5**, 3203 (2014).
[60] M. O. Yokosuk et al., *Nonreciprocal Directional Dichroism of a Chiral Magnet in the Visible Range*, npj Quantum Mater. **5**, 20 (2020).
[61] W. Kaiser, S. Sugano, and D. L. Wood, *Splitting of the Emission Lines of Ruby by an External Electric Field*, Phys. Rev. Lett. **6**, 605 (1961).
[62] B. B. Krichevtsov, V. V. Pavlov, and R. V. Pisarev, *Nonreciprocal Optical Effects in Antiferromagnetic* $Cr_2O_3$ *Subjected to Electric and Magnetic Fields*, Zh. Eksp. Teor. Fiz **94**, 284 (1988).
[63] T. Arima, *Magneto-Electric Optics in Non-Centrosymmetric Ferromagnets*, J. Phys. Condens. Matter **20**, 434211 (2008).